# ESSAYS ON ECLIPSES, TRANSITS AND OCCULTATIONS AS TEACHING TOOLS IN THE INTRODUCTORY ASTRONOMY COLLEGE COURSE

*Noella L. Dcruz[1]*

**Abstract:** We occasionally include projects in our learner-centered introductory astronomy college course to enable non-science major students explore some astronomical concepts in more detail than otherwise. Such projects also highlight ongoing or upcoming astronomical events. We hope that students will feel more interested in astronomy through projects tied to astronomical events. In Spring 2012, we offered short essays focused on eclipses, transits and occultations to promote the rare transit of Venus that occurred on June 5th, 2012. We asked students to write two short essays from three that were offered. The essays contained descriptive and conceptual parts. They were meant to serve as teaching tools. 62% of 106 essays from 55 students earned A, B or C grades. 21% of 47 feedback survey respondents felt the essays increased their interest in astronomy. 49% of respondents felt that the essays were not educationally beneficial and should not be offered again. The most common written response to our survey indicated that students need more guidance and better preparation in writing successful essays. Since students found the conceptual parts of the essays difficult, in the future we will provide relevant activities prior to essay deadlines to help students create successful essays.

## ENSAIOS SOBRE ECLIPSES, TRÂNSITOS E OCULTAÇÕES COMO FERRAMENTAS DE ENSINO EM UM CURSO UNIVERSITÁRIO INTRODUTÓRIO DE ASTRONOMIA

**Resumo:** Nós ocasionalmente incluímos projetos em nosso curso universitário introdutório centrado no aluno para permitir aos estudantes que pertencem às carreiras não científicas explorar alguns conceitos astronômicos em mais detalhes do que o normal. Tais projetos também enfatizam eventos astronômicos em curso ou futuros. Esperamos que os alunos se sintam mais interessados na astronomia através de projetos ligados a eventos astronômicos. No termo de Primavera de 2012 (EUA), propomos ensaios curtos focados em eclipses, trânsitos e ocultações para promover o raro trânsito de Vênus que ocorreu no dia 5 de junho de 2012. Pedimos aos alunos que escrevessem dois ensaios curtos dentre três que foram propostos. Os ensaios continham partes descritivas e conceituais. Eles foram feitos para servir como ferramentas de ensino. 62% de 106 ensaios de 55 alunos ganhou graus A, B ou C. 21% dos 47 entrevistados que responderam ao levantamento posterior sentiu que os ensaios aumentaram seu interesse na astronomia. 49% dos inquiridos consideraram que os ensaios não eram benéficos para a educação e que não devem ser propostos novamente. As respostas escritas da nossa pesquisa indicaram que os alunos precisam de mais orientação e melhor preparação para escrever ensaios bem sucedidos. Como os alunos consideraram difíceis os aspectos conceituais dos ensaios, no futuro iremos fornecer atividades relevantes antes dos ensaios para ajudar os alunos a escrevê-los com sucesso.

**Palavras-chave:** Carreiras não científicas universitárias; Investigação sobre ensino/aprendizagem; Abordagens de ensino; Ensaios; Sistema Solar; Trânsitos.

---

[1] Department of Natural Sciences, Joliet Junior College. E-mail: <ndcruz@jjc.edu>.

# ENSAYOS SOBRE ECLIPSES, TRÁNSITOS Y OCULTACIONES COMO HERRAMIENTAS DE ENSEÑANZA EN EL CURSO UNIVERSITARIO INTRODUCTORIO A LA ASTRONOMÍA

**Resumen:** Ocasionalmente, incluimos proyectos en nuestro curso de introducción a la astronomía universitario centrado en el alumno para permitir que los estudiantes de carreras no científicas exploren algunos conceptos astronómicos en más detalle que lo habitual. Estos proyectos también ponen en relevancia eventos astronómicos en curso o futuros. Esperamos que los estudiantes se sientan más interesados en la astronomía a través de proyectos vinculados a eventos astronómicos. En el período de primavera de 2012 (EUA), propusimos breves ensayos centrados en los eclipses, tránsitos y ocultaciones para promover el raro tránsito de Venus que se produjo el 5 de junio de 2012. Le pedimos a los estudiantes que escribieran dos ensayos cortos de tres que se proponían. Los ensayos contenían partes descriptivas y conceptuales. Los mismos estaban destinados a servir como herramientas de enseñanza. 62% de los 106 ensayos de 55 estudiantes obtuvo grados A, B o C. 21% de los 47 encuestados que respondieron al cuestionario posterior consideró que los ensayos aumentaron su interés por la astronomía. 49% de los encuestados consideró que los ensayos no eran educacionalmente útiles y que no deben ser propuestos de nuevo. Las respuestas escritas más comunes a nuestra encuesta indicaron que los estudiantes necesitan más orientación y una mejor preparación en la redacción de ensayos exitosos. Dado que los estudiantes encontraron las piezas conceptuales de los ensayos difíciles, en el futuro vamos a ofrecer actividades pertinentes antes de los plazos de redacción para ayudar a los estudiantes a crear ensayos de mayor calidad.

**Palabras clave:** Carreras universitarias no científicas; Investigación en la enseñanza/aprendizaje, Enfoques de enseñanza; Ensayos; Sistema Solar; Tránsitos.

## 1.    Introduction

Joliet Junior College (JJC), a community college in Joliet, Illinois, USA, offers an introductory astronomy course aimed at non-science majors. Our introductory astronomy course involves a survey of astronomy in one semester via lecture only. In order to sign up for the introductory astronomy class, students need to be able to do basic arithmetic and to be able to be competent enough in English that they can enroll in the first course in college level English writing at JJC. Students who take this course have a wide spread in age, ranging from traditional age college students (18 – 22 years old) to older students. This course is one of several that students can choose from in order to fulfill their physical science requirement for their Associate's degree. (Community colleges in USA typically offer associate degrees, after completion of two years of a specific set of college courses. Students complete their Bachelor's degree at another institution.) Students can choose courses in other disciplines such as chemistry, physics, geography and geology besides astronomy to fulfill the physical science course requirement.

JJC offers the introductory astronomy course in face-to-face and online formats. In this paper, we are concerned with only the face-to-face format. We use a variety of teaching strategies to keep students engaged during class such as lecture-tutorials, think-pair-share questions, small group discussions and ranking tasks to keep students engaged during class. These have been proven to help students improve their understanding and mastery of astronomy concepts (SLATER; ADAMS, 2003). Outside of class, students do online quizzes and homework that mostly involves ranking tasks to further their learning.

Occasionally, we include projects in our astronomy course, so students can explore certain topics and associated concepts in more detail. In the past, we have offered group poster projects. Our students were not always able to meet with their group outside of class, and they felt this was the main concern with group projects (D'CRUZ, 2009). Because of this, in Spring 2012, we chose to offer individual short essays. Our essays were designed to have both descriptive and conceptual portions, and were intended to be teaching tools rather than projects that mainly involved gathering and presenting information.

We usually structure our projects around sky events and astronomy events that are in the news because students and the public tend to be more interested in the promoted topics. For example, total solar eclipses capture the attention of people all around the world, not just in the zone of totality (PASACHOFF, 2009; PASACHOFF, 2010). Both solar and lunar eclipses capture the interests of the general public and students because such events can be observed easily and they are highlighted in the news (BENNETT, *et al.*, 2012; LITTMANN; ESPENAK; WILLCOX, 2008). An astronomy event that was in the news a few years ago was the purported 400[th] anniversary of the invention of the telescope in 2008. We offered group projects focused on telescopes to commemorate this anniversary, in Spring 2008 (D'CRUZ, 2009).

In Spring 2012, we decided to focus our two sections of introductory astronomy around eclipses (which we always cover) and two other similar astronomical events called transits and occultations. The motivation for this was the rare event, "the transit of Venus," that occurred on 5[th] June 2012 for the USA (ESPENAK, 2011; also available at <http://eclipse.gsfc.nasa.gov/OH/transit12.html>; PASACHOFF, 2012). A transit occurs when an object that looks small on the sky passes in front of an object that looks larger on the sky (WINN, 2011). The smaller object appears as a black disk against the face of the larger object. It occurs when the Earth (or the observer) and two other astronomical objects happen to be in a straight line as they move in space. For example, an annular solar eclipse seen from the Earth can also be called the transit of the Moon across the Sun's disk. When an observer sees an apparently larger object obscure an apparently smaller one, the event is called an occultation (WINN, 2011; <http://www.occultations.org/>).

The previous transit of Venus occurred on Jun 8[th] 2004 (universal time), and the next one will occur on 11 Dec 2117 (ESPENAK, 2011), hence this is indeed a rare event. It is rare because Venus' orbit is inclined by 3.4 degrees to the Earth's orbital plane (ESPENAK, 2011). Venus travels through the Earth's orbital plane in early June and December, but the Earth does not line up with the Sun and Venus at those times because of the differing speeds with which Earth and Venus travel in their orbits around the Sun. The combination of the non-zero inclination of Venus' orbit and the differing orbital speeds of Earth and Venus causes the transit of Venus to recur with gaps of 8 years, 105.5 years, 8 years and 121.5 years currently (ESPENAK, 2011). Because the next transit of Venus will not occur in our lifetime, we felt it was essential to bring this exciting and rare event to the attention of our students in Spring 2012. The event could also be viewed from North America, weather permitting, so students could witness it for themselves if they wished.

The astrophysical significance of the transit of Venus is historical − it was used to measure the distance between the Earth and Sun (ESPENAK, 2011; PASACHOFF, 2012). Currently, radar measurements are far more accurate in determining this

41

distance; hence this transit event appears not be as scientifically important now. Pasachoff (2012) describes some current scientific studies that can be carried out during a transit such as learning about Venus' atmosphere.

Transits occur in other situations too, such as when one of Jupiter's moons passes in front of its disk e.g. http://www.hubblesite.org/newscenter/archive/releases /2004/30/. Another situation that involves transits is in the detection of planets orbiting other stars (e.g. the Kepler mission, BORUCKI, 2010; <http://kepler.nasa.gov>). Such planets are called exoplanets or extrasolar planets. Exoplanets and their parent stars are too far from Earth to image the smaller planet passing in front of the larger parent star (which is the transit detection method discussed so far). Instead, the star's brightness is measured frequently, and the transit is detected photometrically when the star's brightness dims slightly as its planet passes in front of it (WINN, 2011). This dimming repeats after a time interval of one orbital period, indicating the presence of a planet. The frequent detection of transits these days, especially in the discovery of exoplanets, was another motivating factor in choosing the course theme.

To enable students to explore the course theme, we covered the topics in lecture, we included theme-related questions on the tests and final exam and we asked students to write two theme-related essays. Essays and writing assignments that focus on astronomy concepts are used in introductory astronomy classes to enable students to demonstrate mastery of the topics being taught and encourage students to think more deeply about these topics (GREENSTEIN, 2013; SLATER, 2008). Slater (2008) indicates that writing within a particular discipline is important because it helps with learning the content and it helps to develop the ability to communicate successfully within that discipline. Slater (2008) also mentions that students need substantial guidance with writing essays in order to be successful, and that writing assignments that take 5 minutes or less are easier to implement and therefore have been more successful than essays.

We use multiple-choice questions extensively in our class. Having theme-related essays allowed us to gain deeper insight into how students think about the theme because our essays included conceptual parts in addition to descriptive ones. We would not have been able to gather such extensive information about student learning and preconceptions regarding the theme concepts solely through multiple-choice questions.

At the beginning of the Spring 2012 semester, we had asked our students to sign a statement, which said that they allowed their responses to all parts of the class to be used and shared in studies such as this one. The statement also mentioned their responses would be kept confidential. Students were given the option of not signing the statement, and students who chose not to sign have not been included in this study.

This paper is organized as follows: Section 2 covers the astronomy and general education goals of this project. Section 3 explains the essay assignment details. Section 4 provides information on the submitted essays and suggestions for possible future modifications. Section 5 details feedback received from students regarding the essays. Section 6 explains how our students fared on test and final exam questions related to the course theme. Section 7 contains the implications for future essay offerings, and Section 8 contains our conclusions.

## 2. Astronomy and general education goals

Since the sky event we were focused on in Spring 2012 was the transit of Venus, we had three major goals for this project:

- Could students explain what is meant by the "transit of Venus?" (Goal Ia)
- Could students explain is meant by a transit, an occultation and an eclipse via lecture and essays? (Goal IIa)
- Would essays on transits, occultations and eclipses increase students' interest in astronomy? (Goal III)

The first major goal of this project, Goal Ia, is tied to why the transit of Venus occurs rarely. So an associated goal was for students to know why the transit of Venus occurs rarely (Goal Ib).

Since the transit of Venus and transiting exoplanets were the main motivating factors behind the course theme, we decided to have our students learn about some of the factors that affect transits. Students were expected to learn the following in connection with our second major goal:

Goal IIb: what is meant by angular size of a spherical object and how it depends on the object's radius, and distance from Earth.

Goal IIc: predict whether a transit will occur if the rotation period of the transiting object is altered or if the rotation period of the background object is altered, assuming all other parameters stayed unchanged; and to predict whether the duration of the transit will be altered if a transit does occur. The duration of the transit is illustrated in Figure 1.

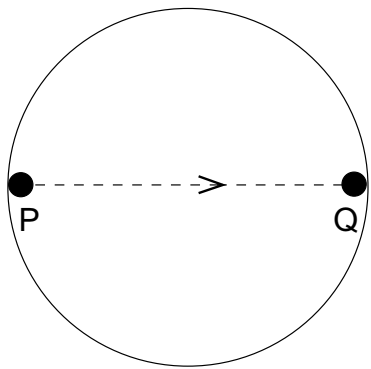

Black dot = transiting object

Large circle = background object

**Figure 1: We define the duration of the transit to be the time taken for the transiting object to travel from point P to point Q of the background object as shown above.**

Goal IId: predict whether a transit will occur if the angular size of the transiting object is altered or if the angular size of the background object is altered, assuming all other parameters stay unchanged; and to predict whether the duration of the transit will be altered if a transit does occur.

Goal IIe: determine how the orientation of the orbit affects whether a transit occurs and how long it lasts if it does occur (see Figure 2).

Original orbit is edge−on and
lines up with larger object's diameter

New orbit is edge−on, but does not
line up with larger object's diameter

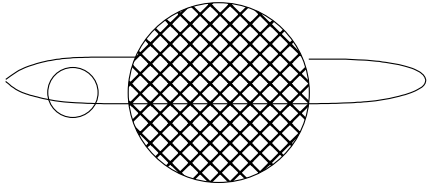 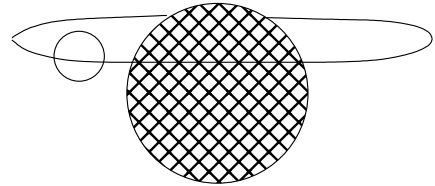

Original orbit is edge−on and
lines up with larger object's diameter

New orbit is
face−on

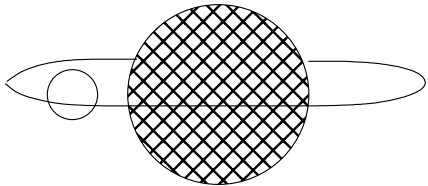 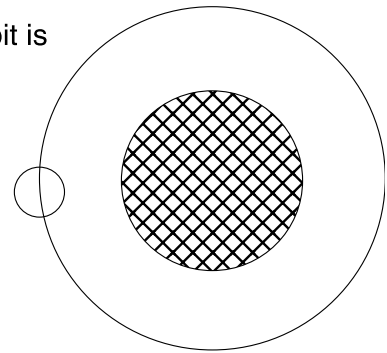

Original orbit is edge−on and
lines up with larger object's diameter

New orbit is
edge−on, at
90 degrees
to original
orbit and
lines up with
larger object's
diameter

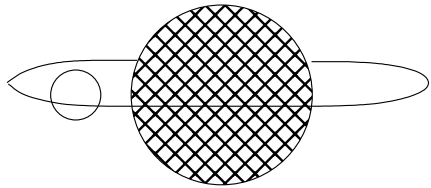 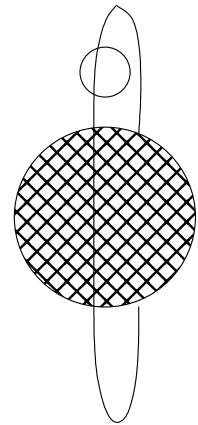

**Figure 2: This figure shows a transiting object's original orbit on the left, and three altered orbits on the right. The small circle is the transiting object for all panels. The large hatched circle is the background object in all panels except for middle right, where both objects are the same distance from the observer. An edge-on orbit refers to one seen in side-view; a face-on orbit refers to one seen from above. Students were asked to predict if a transit would still occur for one of the altered orbits in Essay 1.**

Goal IIf: predict whether changing a transiting exoplanet's rotation period, orbital radius or orbital period will alter the duration of the transit.

Goal IIg: predict whether changing a transiting exoplanet's rotation period, orbital radius or orbital period will alter the amount dimming of its parent star when the planet is completely in front of the star.

We aimed to achieve goal I through lecture, and goal II through background information provided during lecture and through essay assignments. We informed students that besides the essay assignments, a few test questions would be related to these goals in addition to our usual slate of questions on eclipses and exoplanets.

In order to predict how successful our students would be in correctly knowing and explaining the terms listed in Goals Ia and IIa, we relied on results from Prather *et al.* (2005). These authors studied the efficacy of learning in the introductory astronomy class via lecture only or lecture-tutorials. A lecture-tutorial is a structured, conceptual, workbook activity, that students complete during class and therefore can receive immediate feedback from their instructor. Prather, et al., find that 52% of questions in a 68-question survey are answered correct when students are taught solely through lecture, and that 72% of answers are correct when lecture is shortened and combined with in-class lecture tutorials. Since our students learned about Goal 1a through lecture and about Goal IIa through lecture, think-pair-share questions and out-of-class essays, we felt that if students scored at least 52% when explaining these two goals, we would be successful. Lightman and Sadler (1993) show that teachers tend to vastly overestimate the learning achieved by their students when they are asked to make similar predictions. This is why we tied our prediction to a robust study of learning in the introductory astronomy class, but we cannot tell how valid this cut-off is for this paper. There is no similar study on the efficacy of learning through essays in the introductory astronomy class as mentioned in Section 1. Therefore, we are unable to predict how well students would fare on Goal IIa through essays alone.

The third major goal of the project was to increase student interest in astronomy via essays. We were unable to predict the outcome of this goal because it is usually not possible to predict attitudes. But we obviously wanted a majority of students to have an increased interest in astronomy.

Goals I and II, and the standard goals of the introductory astronomy class (SLATER; ADAMS, 2003), help to develop students' logical and critical thinking skills. Critical thinking is one of the institutional general education goals of a science class for non-science majors at JJC. Essay assignments enable students to communicate via writing, which is another institutional general education goal of a class like this one. We do not assess these two general education goals specifically in this project, but students were able to practice the associated skills through the essays and conceptual questions outlined here. (General education goals at the college level in the US typically refer to skills associated with oral and written communication, quantitative literacy, scientific literacy and critical thinking. Some of these skills are common to multiple disciplines and are used frequently outside of formal educational settings. Hence, they are emphasized at the Associate's and Bachelor's degree levels in the US.)

Individual essay assignments had additional goals that are detailed in Section 3.

### 3. Details of Essay Assignments

Students were asked to submit two of the three essays described below. This allowed them some choice in essay topics and deadlines, and staggered our grading. We have found that assigning students at least two essays or two projects usually helps students develop the skills needed to gather and organize information for such assignments (D'CRUZ, 2009; SLATER, 2009). Since grading individual submissions requires considerable time commitment, we told students we would grade the first two essays they submitted.

We chose to assign individual essays because these do not rely on group effort. We have assigned group projects in the past, and have given students up to 90 minutes of class time per project to facilitate working in a group. However, students still needed to meet outside of class to complete their project, and it was difficult for many groups to find a common meeting time (D'CRUZ, 2009; D'CRUZ, 2010).

We also chose to assign short essays so that a small part of the course grade would be devoted to essays. These were not meant to be high stakes assignments. We actively encouraged all students to submit essays as related questions would appear on tests and the final exam.

Students were expected to submit essays that were 3 − 5 pages long, with double-spaced 12 pt font text, inclusive of bibliography (around 1200 − 1500 words). An additional page or two could be used for images. Each essay was worth 2.5% of the course grade. All essays had grading rubrics associated with them and detailed guidelines of what was expected of students. Students were given two weeks to complete an essay.

Students were informed that all essays would be checked for originality using turnitin, an anti-plagiarism service that JJC subscribes to (http://www.turnitin.com). They were asked to submit electronic versions of their essays directly to the turnitin dropbox available through the college's online learning management system.

The English pre-requisite for this course requires that students should have sufficient English skills to be able to sign up for the first English writing class required for an associate in arts and associate in science degrees. (JJC does not offer Bachelor's degrees.) We took this into consideration when designing the assignments. These assignments were not expected to produce research papers, however we did expect students to have a well-written introduction and conclusion, to break up their writing into paragraphs that flowed smoothly and to use correct grammar, spelling and punctuation. About one-third of each essay's grade was devoted to these basic English skills. Appendix A explains how the grading of English skills was separated from the grading of content.

Students were expected to provide a bibliography of their sources in the Modern Language Association (MLA) format, and to cite their sources within their essay. To assist students with this we directed them to Purdue University's Online Writing Lab guidelines on the MLA format (http://owl.english.purdue.edu/owl/resource/747/01/). When grading, we focused only on the bibliography, and assigned it full points if all sources were listed in sufficient detail whether or not they were in MLA format. We did not penalize students for not citing sources within their essays.

We were unable to grade drafts because there was insufficient time during the term to grade and return drafts while offering three essays. But we offered to read Essays 1 and 2 prior to submission so students could know whether or not their essay needed further work. Our handouts informed students that since no grade was being assigned at this time, there was no guarantee of a high grade if we said that their essay draft was good. We also did not check for plagiarism at this reading. Two of our students took advantage of this, and we feel this helped them to improve their essays.

However, during through the Spring 2012 semester, we encountered a student in our Life in the Universe class, where similar short essays were offered, who was displeased with this approach. We had told him that his essay was fine prior to submission, but we found the essay to be substantially plagiarized when we graded it, and therefore assigned it zero points. As a result, the student complained to our department chair and dean that our comments on his draft were misleading. Because of this student's complaint, we did not offer to read Essay 3 drafts from our introductory astronomy students, though we answered specific questions about the essay.

We also directed students to the college's writing and reading center for help with their essays. We did not ask students if they used this service.

### 3.1 Description of essays

### 3.1.1 Essay 1: Transits and Occultations involving solar system objects

The first essay was offered immediately after solar and lunar eclipses and angular size were covered in lecture. This assignment enabled students to become familiar with the terms transit, occultation, and angular size (goals IIa, b). For this essay, students were asked to describe a transit and an occultation event, excluding solar and lunar eclipses and transiting exoplanets. They were asked to search the "Astronomy Picture of the Day" website for these events and their descriptions (http://apod.nasa.gov) and to include photographs of the events in their essay. They could use other websites if they wished. We provided links to examples of each event from Astronomy Picture of the Day within the college's online learning management system, and explained how students could use the search feature on the Astronomy Picture of the Day website to search for transits and occultations.

The conceptual portion of the essay was twofold. Students were asked to predict whether a transit would still occur and whether its duration would be longer, shorter or remain unchanged if the angular size or rotation periods of the foreground and background objects were changed while everything else stayed unchanged (they needed to address only one of these four). The duration of the transit was defined as shown in Figure 1. They were also asked to predict whether a transit would still occur and whether its duration would be longer, shorter or remain unchanged if the orbital path of the apparently smaller object were altered as shown in Figure 2 (they needed to address only one of three options shown). These two "stand-alone" conceptual questions were designed to help them to master the geometry of transits and are related to goals IIa,b,c,d,e.

We also asked students to write about what might have not been clear to them (so we could help them with improving their understanding of these topics) and what more they would want to know about such events. This, along with the predictions, provided some uniqueness to submissions, which reduced the chance of plagiarized essays.

### 3.1.2 Essay 2: Spacecraft and Transits, Occultations and Eclipses

The second essay was offered after gravity, orbits and the electromagnetic spectrum were covered. This assignment asked students to describe a transit, occultation or eclipse event that was recorded by a spacecraft. Their event had to include at least one solar system object besides the spacecraft i.e. they could not choose a transiting exoplanet for this essay. They were asked to state in which part of the electromagnetic spectrum the event was recorded as part of their description.

Students were also asked to describe the scientific mission of the spacecraft, its orbital path and how it was maneuvered into its orbital path. The goals of this assignment involved goal IIa from Section 2 above (including identifying the three objects that were aligned in their event), learning about a spacecraft mission, the region of the electromagnetic spectrum that the spacecraft operates in and how a spacecraft is launched and directed to its destination. We explained how orbital paths are altered when this essay was given to students, and we directed them to the part of our textbook where this is covered (BENNETT *et al.*, 2012). We also asked them to include concepts associated with gravity such as gravitational force, acceleration, speed, velocity, etc. in this part of the essay to demonstrate their understanding of these concepts. This last part formed the conceptual portion of Essay 2.

We asked students to write about what may have been unclear to them about the science of their spacecraft and achieving the orbit of their spacecraft and/or what more they might wish to know about achieving the orbit of their spacecraft. This was done to deter plagiarism and for us to be aware of what difficulties students may have had.

We provided students with a list of transits, occultations and eclipses recorded by spacecraft, so students could be sure they picked an appropriate event for their essay. We provided links to the spacecraft websites, but allowed the students to search for information on the events on their own. This information was available through JJC's online learning management system.

### 3.1.3 Essay 3: Discovering Exoplanets using the Transit Method

The third essay assignment involved transiting exoplanets, and was offered after this topic was covered in lecture. Transiting exoplanets cannot be imaged like the events that Essays 1 and 2 involved, because the planet and its star are too far from Earth. Instead they are detected by the dip in the star's brightness when the planet is in front of the star (WINN, 2011). Since students had been introduced to transits during lecture and by doing at least one of the first two essays, we hoped they would feel comfortable exploring transiting exoplanets even though it is impossible with current instruments to obtain a photograph of the event.

Students were told to imagine that they had just discovered a transiting exoplanet and were excited to tell everyone about its discovery and properties. We asked students to choose a transiting exoplanet from the Exoplanet Encyclopedia (http://www.exoplanet.eu). In order to assist students, we provided a link to the Exoplanet Encyclopedia in JJC's learning management system and we showed students how to choose transiting exoplanets from this database.

Students were expected to briefly describe the experiment responsible for their planet's discovery and to identify the part of the electromagnetic spectrum that the experiment used in the discovery. We directed students to the list of exoplanet experiment websites maintained by the Exoplanet Encyclopedia website, so they could find the pertinent information for their exoplanet.

The assignment asked them to indicate whether the exoplanet is in the Milky Way galaxy or outside of it. They also had to include the exoplanet's orbital and physical properties; a table of this information was sufficient. They were asked to compare the shape and size of their exoplanet's orbit to the orbit of a solar system planet, so that the reader could have some insight into the exoplanet's orbital size and shape. This gave students a chance to revise Kepler's first and second laws of orbital motion, and properties of an ellipse such as eccentricity, focus, variation of orbital speed within an elliptical orbit, etc. (BENNETT *et al.*, 2012). The assignment asked whether the chosen exoplanet was jovian (Jupiter-like) or terrestrial (Earth-like) or neither, and to explain how they arrived at their choice. Properties of jovian and terrestrial planets had been discussed in detail when we covered our solar system during lecture (BENNETT *et al.*, 2012). This was an opportunity for students to explore these types of planets further.

In order to help students grasp the transit technique in the context of exoplanets, we asked students to predict how the transit duration and the parent star's brightness during the transit would change if the orbital period, orbital radius or rotation period of the planet were changed (they needed to address only one of the latter three). These questions were related to goals IIb,e,g, and Kepler's third law of orbital motion (BENNETT *et al.*, 2012). They were also asked to write about what more they would like to know about their exoplanet, its star and its discovery and what may not have been clear to them about the discovery of the exoplanet. This last part, together with the conceptual questions, also served to reduce the chance of plagiarized submissions.

## 3.2 Grading of Essays

Grading of each required item in an essay was done on a very good, satisfactory, unsatisfactory basis, usually using a 3, 2, 1 point scale. With this scale, it was not possible to earn a zero for it unless the essay was substantially plagiarized or a student did not turn in an essay. Also, if an essay contained one or two minor errors, it was possible for the student to earn 100% with this grading scale. As mentioned above, about one-third of the points were for appropriate English language usage including writing a good introduction and conclusion using the same scale. Detailed rubrics are available on request.

## 4.    Overview of essay submissions and possible future modifications

63 students completed the course. We received 106 essays from 55 students. Of these, 23 were Essay 1 submissions, 37 were Essay 2 submissions and 46 were Essay 3 submissions. 8 students did not submit essays, and 4 submitted one essay only.

The grade distribution of the average essay score (the average of two essay scores) was as follows: A = 21.7%, B = 21.7%, C = 19%, D = 15% and F = 22.6%. The distribution of letter grades for each essay is shown in Table 1. The course grade distribution is shown in Table 2 for students who did each essay and for all students who completed the course. For all of these distributions we used the following grade scale: A = 88% to 100%; B = 78 to 87.99%; C = 68 to 77.99%; D = 58 to 67.99%; F < 58%. We elaborate further on these distributions in Section 4.5 below.

| **Essay Letter Grade** | Essay 1 (n=23) | Essay 2 (n=37) | Essay 3 (n=46) |
|---|---|---|---|
| **A** (88% to 100%) | 56.5% | 16.2% | 17% |
| **B** (78 to 87.99%) | 13% | 32.4% | 9% |
| **C** (68 to 77.99%) | 21.7% | 13.5% | 24% |
| **D** (58 to 67.99%) | 8.7% | 16.2% | 17% |
| **F** (< 58%) | 0% | 21.6% | 33% |

**Table 1: Percentage of students scoring A, B, C, D, F grades for Essays 1, 2 and 3. The number of students who submitted each essay is shown in brackets in row 1. The conversion of percentages to letter grades is shown in brackets in the leftmost column and at the beginning of Section 4. The percentages in each column may not add up to 100% because of rounding.**

| **Course Letter Grade** | Course grade for Essay 1 submitters (n=23) | Course grade for Essay 2 submitters (n=37) | Course grade for Essay 3 submitters (n=46) | Course grade for all students (n=63) |
|---|---|---|---|---|
| **A** | 26.1% | 13.5% | 10.9% | 12.7% |
| **B** | 26.1% | 21.6% | 21.7% | 20.6% |
| **C** | 30.4% | 35.1% | 37.0% | 36.5% |
| **D** | 17.4% | 27.0% | 26.1% | 23.8% |
| **F** | 0% | 2.7% | 4.3% | 6.3% |

**Table 2: The distribution of course grades for students who did Essays 1, 2 and 3 and for all students who completed the course are shown as percentages. The number of students involved in each column is shown in brackets. The percentages in each column may not add to 100% because of rounding. The letter grade ranges are given in Table 1 and at the beginning of Section 4.**

Two essays were substantially plagiarized i.e. over 60% of the essay content and bibliography matched sources according to turnitin.com, and a third essay was 48% plagiarized. The two students involved were warned that their first essay submissions violated JJC's Academic Honor Code. The substantially plagiarized essay earned a zero, while the 48% plagiarized essay was penalized 48%. The students were informed that a second plagiarized submission would be reported to the college. They were also given a copy of the turnitin.com match so they were aware of why their essay was considered plagiarized. One of these two students submitted a second essay that was substantially plagiarized, and was reported to the college for violating JJC's Academic Honor Code. This essay was also given a score of zero.

We note that the English language writing skills (grammar, spelling, punctuation, flow, formatting, introduction and conclusion, etc.) of our students averaged 85% in Essay 1, 73% in Essay 2 and 70% in Essay 3. While the scores for Essays 2 and 3 are satisfactory, these could be improved. However, within the scope of this course, it would not be possible for us to spend class time or provide detailed guidance on this.

All essays contained descriptive and conceptual portions. The descriptive parts had concepts embedded in them, but for the details presented below and in the rest of the paper, when we mention conceptual questions, we are referring to the "stand-alone" conceptual questions in Essays 1 and 3 detailed above, not the ones that were part of the descriptive portion. We also refer to writing about gravity related concepts in Essay 2 as the "conceptual" part of this essay.

### 4.1 Review of Essay 1 submissions

Essay 1 met with the greatest success. The average score was 86% for this essay. 22% of the essay score was for conceptual questions, and students averaged 75% on this. Students did extremely well on the descriptive part of the essay (Goal IIa dominated this part), averaging 90%, indicating that they felt very confident explaining what is a transit and an occultation.

91% of students (21 of 23) were able to easily understand how the orientation of the orbit affects whether a transit occurs or not (see Figure 2 and goal IIe), though they did not all address how the transit duration would change. 55% of those who did were correct (6 of 11).

The effect of rotation period or angular size on the transit and its duration (goals IIc, IId) had very few correct responses. 14% of responses (3 of 22) correctly indicated that rotation period does not affect the transit or its duration, and two of these three students included excellent descriptions of how to distinguish rotation from orbital motion. 27% of responses (6 of 22) addressed how the angular size of either object affects the transit and all were correct. Half of these students correctly explained how the transit duration changes with angular size; the other half did not address this part. 27% of responses (6 of 22) showed that students were unable to distinguish between rotational and orbital motion, and therefore were not able to respond correctly. We had not anticipated this; otherwise we would have provided some pre-emptive guidance ahead of time. 27% of responses were unclear (6 of 22). 4.5% of responses (1 of 22) attempted to incorrectly use the angle between Earth's orbit and the transiting object's orbit as the angular size of the transiting object. One student, out of the 23 who did this essay, did not address this question.

In the future, in addition to explaining the difference between rotational and orbital motions, we could have an in-class activity covering the conceptual questions of Essay 1 prior to the deadline. We could follow this up with an online quiz administered through the college's online learning management system that closes at least a week before the essay deadline. This would allow students to practice and develop confidence with these concepts before turning in the essay.

## 4.2 Review of Essay 2 submissions

The average score for Essay 2 was 70%. Students had considerable difficulty with including concepts related to gravitational force, acceleration, speed and velocity in their essay. This formed 19% of the essay score, and students averaged 57% for this, mainly because several of them did not include some examples of gravity or speed in their essays. They were able to do well on the rest of the essay and their scores averaged 74% here. Students averaged 73% when describing their event i.e. the portion of the essay that involved Goal IIa. Hence Goal IIa met with success in Essay 2.

We covered gravity concepts through lecture, accompanied by existing activities (lecture tutorial, think-pair-share questions, ranking tasks). However, it is clear that this coverage is not sufficient in helping students to write about the concepts. Here is an example of what we expect if a student chose the Cassini spacecraft: "The gravitational force between Earth and the Cassini spacecraft when 10 km separates them is smaller than between Saturn and Cassini when they are 10 km apart because Earth is lighter than Saturn."

In the future, we will spend some class time helping students demonstrate their understanding of these concepts through writing. We could provide students with examples. Then students could draft what they plan to include in their essay. We could spend 60 minutes of class time, divided over two classes to review these drafts so students are appropriately prepared to address this part of the essay. Helping students with writing scientific concepts could help them to more deeply understand and retain these concepts, so it would be worthwhile to spend class time on this (GREENSTEIN, 2013; SLATER, 2008).

## 4.3 Review of Essay 3 submissions

The average score for Essay 3 was 67%. Students averaged 74% on the descriptive parts of the essays, which is good. They averaged 70% on the part that included Goal IIa, indicating that this goal met with success. On concepts related to Kepler's laws, jovian and terrestrial planets and transits, students averaged 56%. These conceptual questions made up 34% of the essay grade.

Around half the students addressed how the shape and size of their planet's orbit compared to that of a solar system planet (Kepler's 1st Law for shape and size, and Kepler's 2nd Law for shape). Most of these gave excellent comparisons. 72% (33 of 46) addressed whether or not their planets is terrestrial or jovian. About 76% (25 of 33) of these students could correctly classify their planet as terrestrial or jovian, and around half of these provided compelling explanations for their choice using the size and masses of their planets.

The above numbers indicate that students need some help in order to be more successful with these two parts of the essay. Perhaps they could be given examples of what would be suitable to write and/or some prior practice with this. Alternatively, this part could be re-worded so that the student is asked to imagine visiting their exoplanet.

Then they can describe what they experience as the planet rotates and orbits, whether they experience seasons, whether they could stand on the surface of the planet, whether their planet has an atmosphere and whether they could withstand the temperature at the surface of the planet. As long as they correctly addressed at least three of these very well and compare them to the Earth or a solar system planet of their choosing, they would receive full points. Such an approach might appeal more to students and could help them with turning in better essays. It would also serve to keep the chance of plagiarized submissions low.

With the transit duration related question in Essay 3, there were 48 responses from 41 students; 4 students addressed more than one of the options presented and 5 did not address this. 66% of responses (35 of 48) for this part of the essay were correct, which is very encouraging. 62.5% of responses (30 of 48) addressed how the transit duration is affected by increasing the orbital period (Goal IIf). 73% of these answers (22 of 30) were correct – the transit duration increases because the orbital speed decreases when the orbital period increases according to Kepler's 3rd Law. All responses explaining the effect a decrease in orbital radius has on transit duration (17% or 8 of 48) were correct – the transit takes less time because the planet orbits faster when closer to the star according to Kepler's 3rd Law (Goal IIf). 12.5% of responses (6 of 48) involved the option related to the planet's rotation period (Goal IIf). 83% (5 of 6) of these correctly answered that increasing the rotation period of the planet does not affect transit duration. The remaining 8% of responses (4 of 48) contained answers that were unclear.

When addressing what affects the dimming of the star during the transit for the options provided, 39 students provided one response each, while 15% of students who did this essay (7 of 46) did not attempt this question. 36% of the responses were correct (14 of 39), which is disappointing (Goal IIg). 13% of responses (5 of 39) were unclear. 23% of responses (9 of 39) addressed the increase in rotation period of the planet, and all of these students correctly noted that rotation period does not affect the dimming of the star. The other two options made up 64% of responses (25 of 39). 21% of responses related to the effect of decreasing orbital radius on the star's dimming during a transit, (3 of 14), correctly explained that the star's brightness drops less when the planet's orbital radius is decreased because the planet's angular size decreases for the observer. 18% of responses involving the effect of an increase in orbital period on stellar dimming during a transit, (2 of 11), correctly indicated that the star's brightness is dimmed more when the planet's orbital period increases because the planet's angular size increases for the observer.

This last question is possibly the least successful of the conceptual questions. 21% of responses (8 of 39) indicated that the question was interpreted as asking for how much time the star's dimming is affected, not how the intensity of the dimming changes, despite having a figure included in the handout to accompany this question. Such an interpretation allowed students to arrive at the correct answer for the wrong reasons − something that would not have been revealed if this were presented as a multiple-choice question. Perhaps this question should be removed from the essay assignment and offered as an in class activity, and/or re-worded.

As with the transit related concepts in Essay 1, if we provided students with practice with the concepts in Essay 3 ahead of time, it is likely that more of them would be able to arrive at the correct conclusions to include in their essays. We could provide examples in the essay handout and/or we can have students do similar questions in class so students know what is expected of them. We could offer quizzes administered through the course learning management system at least a week prior to the essay deadline to follow up on activities done in class. We could also couple these with reducing the number of conceptual questions in the essays (e.g. remove the one about the dimming of the star in Essay 3). Given that other course material needs to be covered during the semester, it may not be possible to devote class time to all of the concepts we have included in Essays 1 and 3, and we would adjust the essays accordingly.

### 4.4 What did students learn from the essays?

From these essays, we find that students successfully learned about what is a transit, what is an eclipse and what is an occultation, and they recognized that two objects and the Earth (or a spacecraft) need to be aligned in order to observe such events (Goal IIa). They also understood how the orientation of the orbit on the sky affects the occurrence of transits (Goal IIe).

Around a quarter of the students were unable to distinguish between rotational motion and orbital motion in Essay 1, though not all students chose to address the concept of rotation period, so this is a lower limit. After Essay 1 submissions were returned to students, we explained to all students the difference in these two motions with demonstrations, including asking students to stand and spin in place to emphasize rotation. It is encouraging to note that this confusion was reduced to 2% in Essay 3, though again this is a lower limit since not all students chose to address the effects of rotation period on transit duration and decrease in stellar brightness during the transit.

### 4.5 Determining the difficulty level of the essays

Table 1 shows the break down of essay grades for the three essays, and Table 2 shows the distribution of course grades for students who did each essay as well as the overall course grade distribution. The course grade distributions for students doing Essays 2 and 3 are similar, and these are similar to the course grade distribution of all 63 students who completed the course. The course grade distribution for Essay 1 contains more As and Bs compared to the same distributions of those who did Essays 2 and 3. Therefore, it is not unexpected that Essay 1 has the highest average score. However, we cannot conclude that Essay 1 is the easiest essay based on the essay grade distribution.

Since students found the conceptual portion of each essay the most difficult, we will use the fraction of the essay score allocated to "stand-alone" concepts as the criterion that determines the difficulty level of each essay. 22% of Essay 1's score was assigned to concepts, and 19% of Essay 2's score was assigned to concepts. Since almost the same percentage of both essays was allocated to concepts, we can assume that the difficulty level of these two essays is comparable. The fact that the average score of Essay 1 submitters was much higher than that of Essay 2 submitters can therefore be attributed to the difference in abilities between the groups of students who did these essays. We clearly see this difference in their course grade distribution in

Table 2. There were 8 students who did both essays, and their average scores were comparable, at 87%. However, this is not a large enough sample from which to draw conclusions about how the essays compare in difficulty. We will conclude solely from the fraction of the score allocated to concepts that the two essays are likely to be comparable in difficulty.

Essay 3 had 34% of its score assigned to concepts, which is a lot more than for either of the other essays. This would indicate that Essay 3 was the hardest of the essays, though we did not intentionally make this essay the hardest. We had not anticipated that students would find all of the concepts in this essay so challenging, otherwise we have reduced the number of concepts before offering this essay. The distribution of course grades for students doing Essay 3 is similar to those who did Essay 2. However, Essay 3 scores show that 50% of the scores are As, Bs and Cs, compared to 62% for Essay 2 scores. This is another indication that Essay 3 was more difficult than the others.

In the case of Essays 2 and 3, 29 students did both, and their average scores were 67% for Essay 2 and 66% for Essay 3. This does not agree with the argument that the essays have differing difficulty. But the averages are close because some of the students who did both essays improved from Essay 2 to 3, while others did worse on Essay 3 than Essay 2, and the rest stayed unchanged. We had 14 students who did Essays 1 and 3, and their scores averaged 85% and 72% respectively. This does point to Essay 3 being more difficult, but since the sample size is small, it would not be reliable to draw conclusions from.

Hence, we use the fact that Essay 3 had a larger number of concepts and that fewer students earned As, Bs and Cs in this essay compared to Essay 2 (with both groups of students performing similarly in the course) to conclude that Essay 3 was the hardest of the three essays. We did not intend to make this essay more difficult than the others.

## 5. Students' attitudes towards essays

We constructed a feedback survey on the essay assignments to find out how students felt about them, so that future essay assignments could be better structured. The survey contained five statements, which requested responses of "Yes," "No", or "Maybe." The survey contained a sixth and final question requesting a written response. 47 students responded to the survey, including two who did not submit essays.

We told students that participation in the survey was voluntary, and that their responses would not be tied to their names. We offered a small amount of extra credit for participating as the survey was done outside of class, through the college's online learning management system.

45 out of 55 students who submitted essays provided us with feedback. The course and essay grade distribution of these 45 students are almost identical to the same distributions for all those who submitted essays. This, combined with the fact the sample size is reasonably large, leads us to conclude that the responses received do not have any *a priori* bias.

The results of how students felt about the five statements are shown in Table 3.

| How do you feel about the statements in questions 1 – 5? | Yes | No | Maybe |
|---|---|---|---|
| Q1. The essay handouts contained sufficient details about what your instructor expected. | 66% | 17% | 17% |
| Q2. The essay grading rubrics were fair. | 70.2% | 23.4% | 6.4% |
| Q3. I felt that the essays increased my interest in astronomy. | 21.3% | 57.4% | 21.3% |
| Q4. The educational value of essays is such that I would recommend them in future offerings of this course. | 27.7% | 48.9% | 23.4% |
| Q5. I would like to witness the transit of Venus on 5th June 2012 if the weather and my schedule permit. The transit starts at 5:10pm. The Sun sets before it ends. | 63.8% | 14.9% | 21.3% |

**Table 3: Student responses to the first five questions of the essay feedback survey. Students were asked how they felt about the five statements listed above.**

Two-thirds of students said the handouts provide sufficient information to create their essays, and 70% find the grading rubrics to be fair. This is encouraging.

We feel that our most significant results from these questions is that 21% of students felt the essays increased their interest in astronomy, and that 49% of students felt that the essays were not educationally beneficial to offer in the future. From Table 1, we see that students who did Essay 1 earned a greater fraction of As, Bs and Cs in this essay compared to those who did Essays 2 and 3. We also can tell from comparing Tables 1 and 2, that for Essays 2 and 3, more students scored lower on essays than in other parts of the course, indicating that the essays were more challenging than other parts of the course. Since there were more students who did Essays 2 and 3, it is likely that 28% of students would find the essays educationally beneficial, 21% would feel that the essays increased their interest in astronomy, and a majority would feel the opposite way.

Table 4, shows the average essay grade distribution of our 47 survey respondents both as percentage and raw numbers in the second row. The table also shows the percentage of students responding "Yes," "No" and "Maybe" to questions 3 and 4 of the survey for each essay letter grade.

In the case of question 3, the influence of essays on student interest in astronomy, 21% of students report an increase in interest in astronomy across all grades i.e. students who scored higher in essays did not feel more interested in astronomy than those who scored lower. This probably indicates that it might be difficult to increase interest in astronomy through this particular set of essays, or through essays that emphasize concepts.

Around 40% of those with higher grades find the essays to be educationally beneficial according to Table 4. 22% of those earning D grades find the essays to be educationally beneficial and this drops to zero for the group of students who earned failing essay grades. If more students could earn higher grades for their essays, it is possible that a majority of students would find the essays educationally beneficial and recommend them to be offered to future introductory astronomy students instead of just 28%.

| Average Essay Grade | A | B | C | D | F |
|---|---|---|---|---|---|
| Percentage of survey respondents, with number of respondents in brackets. | 19% (n=9) | 28% (n=13) | 11% (n=5) | 19% (n=9) | 23% (n=11) |
| *Yes, No, Maybe percentages for Q3 of survey,* "I felt that the essays increased my interest in astronomy," *for each essay grade* | *22.2%, 33.3%, 44.4%* | *23%, 38.5%, 38.5%* | *20%, 60%, 20%* | *22%, 78%, 0%* | *18%, 82%, 0%* |
| *Yes, No, Maybe percentages for Q4 of survey,* "The educational value of essays is such that I would recommend them in future offerings of this course," *for each essay grade* | *44.4%, 22.2%, 33.3%* | *38.5%, 46.2%, 15.4%* | *40%, 20%, 40%* | *22.2%, 66.6%, 11.1%* | *0%, 72.7%, 27.2%* |

**Table 4: The distribution of average essay grades for two essays for all survey respondents is shown in the second row as percentages, along with the number of respondents in each grade. The breakdown of Yes, No, Maybe responses for questions 3 and 4 from the essay feedback survey for each essay letter grade are shown as percentages in the third and fourth rows respectively. The percentages in each cell of rows 3 and 4 may not add to 100% because of rounding.**

In Spring 2008, when we asked our students to do two group poster projects on telescopes, a majority of students, around 45%, felt the projects were educationally beneficial and should be offered again. This compares well with how those who earned As, Bs and Cs on essays in Spring 2012 feel. A majority of our Spring 2008 students, around 40%, also felt that the projects increased their interest in astronomy. The projects in Spring 2008 were less conceptually oriented than the ones presented here and they featured a different topic, so at least these two factors would have affected their interest in astronomy (D'CRUZ, 2009). Perhaps working in groups was also beneficial to students versus working individually.

Our survey indicates that 64% of students would have liked to view the 2012 transit of Venus, which is encouraging. When the event neared, we informed students about the opportunity to view the transit at Joliet Junior College. However, since the semester had ended a few weeks earlier, only one student came to the viewing event.

**5.1 Responses to the sixth survey question**

The sixth and last survey question was:

"Q6. What is the most important thing the instructor needs to know before assigning similar essays next semester in order to improve the assignments?"

44 respondents wrote 68 comments for this question. The main trends of the comments are:

A. 28% of the comments indicated that expectations need to be made clearer, there needs to be less requirements, instructions need to be simplified and students need to be prepared better to meet the expectations. 66% of students felt the handouts clearly explained the expectations according to the first question in the survey, which is inconsistent with these comments. However, it is clear that some students needed help in constructing their essays, and that some students found the essays to be more difficult than other parts of the course. In Section 4, we mentioned that we would provide students with suitable activities to prepare them for at least the conceptual parts of the essays in the future. Two comments suggested that example essays be provided so that students have a better idea of expectations. This can also be done in the future. We hope that these alterations will help students to be better prepared to write essays.

B.  10% of the comments indicated that the essays are fine as is and expectations are clear. This is reassuring.

C.  6% of comments indicated that other essay topics should be considered. The particular theme we chose for Spring 2012 resulted in essay topics that were somewhat narrowly focused. We can certainly provide a wider range of topics for future essays.

D.  6% of comments indicated that our grading should be more lenient. We reported in Table 3 that 70% of students found the grading rubrics to be fair, which is at odds with these comments. We do not feel the grading was biased against students as over 62% of essays were assigned A, B, or C grades. However, as mentioned in Section 4, Essay 3 unintentionally turned out to be the hardest of the three, with the other two being comparable. Since more students did Essay 3 than the other two, it is likely that some students would feel the grading was not lenient. A comment related to this concern mentioned that we should remind students of essay requirements so they can check whether their essays contain the necessary pieces prior to submission. This student also wrote that the reminder could reduce the displeasure some felt upon receiving their essay grade. Since the handouts contain the requirements and grading rubrics, providing such a reminder can be easily done in the future.

E.  The remaining 50% of comments covered a wide variety of topics. Perhaps the most relevant of these to future essay offerings were to make each essay worth more than 2.5% of the course grade and to provide more time to complete the essays.

## 6.    Test and final exam questions related to course theme

### 6.1 Test questions related to course theme

We wanted all students to be able to master goals I and II, that were described in Section 2. But since goal I was not included in the essays, and since not all students did all the essays, we felt that the inclusion of a few related questions on tests and final exams would be a reasonable way to assess their mastery of these goals. We put theme related questions on our tests only after returning graded essays and providing students with feedback on them. As a result, we could include four theme-related multiple-choice questions on only the second test, because we had graded and returned only Essay 1 at this time. (Our tests and final exam contain only multiple-choice questions.)

The test questions addressed goals IIa and IIb. We always have questions on solar and lunar eclipses, and on exoplanets on our tests, so we report here only on the additional questions that were connected to the course theme. The questions and the percentage of students who answered them correctly are given below:

1. The Moon's orbital distance from the Earth is slowly increasing. The Moon's angular size in the future will be larger/smaller/unchanged compared to currently. (Goal IIb)
   48% answered this question correctly, stating the angular size will be smaller.

2. Which of the following best describes an occultation? (Goal IIa)
   (a) an object of smaller angular size passes in front of an object of larger angular size
   (b) an object of larger angular size passes in front of an object of smaller angular size
   (c) a lighter object passes in front of a heavier object
   (d) a heavier object passes in front of a lighter object
   67% correctly chose (b)

3. During an annular solar eclipse, the Moon has a smaller angular size than the Sun, and is unable to completely block out the Sun. This type of eclipse could also be called (Goal IIa)
   (a) a transit of the Moon
   (b) an occultation of the Moon
   (c) a transit of the Sun
   (d) an occultation of the Sun
   49% correctly chose (a)

4. What effect or effects would be most significant if the Moon's orbital plane were exactly the same as the Earth's orbital plane? (Goal Ib)
   (a) Solar eclipses would be much rarer.
   (b) Solar eclipses would be much more frequent.
   (c) Solar eclipses would occur with the same frequency as currently.
   88% of students correctly answered (b).

## 6.2. Transit related final exam questions

Since we wished to highlight the transit of Venus during the course, we felt that the final exam should contain questions regarding this topic (goal I). We also included questions related to goal II. The final exam was comprehensive, which meant we were limited to including five multiple-choice questions addressing these items. These were in addition to our regular coverage of eclipses and exoplanets. To help students prepare for this part of the exam, we provided a practice quiz with similar multiple-choice questions through our learning management system. Students could attempt this quiz as many times as they wished before the exam. There were 6 submissions from 4 students for this quiz. 62 students did the final exam, out of 63 who completed the course.

Between 72% and 82% of students answered the following final exam questions correctly (the exact wording of the questions is available upon request):

(a) What is meant by the transit of Venus? (Goal Ia)
(b) How does Venus' angular size change as its distance from Earth decreases? (Goal IIb)
(c) How does the time taken for Jupiter's moon, Io, to transit across Jupiter's diameter change when Jupiter's rotation period is increased, assuming everything else stays unchanged? (Goal IIc)
(d) If Io's orbital radius increased while Io is centered on Jupiter, will it block more/less/the same amount of Jupiter's disk, assuming everything else stays unchanged? (Io is a moon of Jupiter.) (Goal IIg)

52% of students were able to correctly answer why the transit of Venus occurs so rarely (Goals Ib). The question focused on the fact that Venus' orbit is inclined to the Earth's orbital plane, not on the relative orbital speeds of Earth and Venus (we did not emphasize the latter much in our class). Since students had done very well on the test question concerning the effect of the inclination of the Moon's orbit on the frequency of solar eclipses, we had hoped that they would extend that knowledge to explain the rarity of the transit of Venus. However, this was not the case.

The test and final exam questions showed that students were easily able to comprehend what is meant by a transit and occultation and that their understanding of how the angular size of an object depends on distance improved over the course of the semester.

## 7.    Implications for future essay offerings

Sections 4 and 5 have shown that writing about concepts at a level that demonstrates deep conceptual learning is difficult. The stand-alone transit concepts included in the essays required more guidance and practice than was provided through our brief lectures in order for students to master them. We will provide suitable activities in the future to help students feel more comfortable and confident with these concepts when constructing their essays. Concepts like Kepler's Laws and gravitational force were covered through lecture tutorials, think-pair-share questions, and ranking tasks. When students were asked to demonstrate their understanding of these through writing, there were several who were not ready for these, as most of our usual assessments do not involve writing in such a manner. In the future, we would offer students more guidance and practice with writing about scientific concepts.

Based on the results presented, we feel that we can suitably modify the essay assignments so more future students find them to be educationally beneficial and more can be persuaded to increase their interest in astronomy through the essays. We feel that while essays can be challenging, offering such writing assignments helps students to think about astronomy topics and concepts more deeply (GREENSTEIN, 2013; SLATER, 2008) and they complement the other learner-centered activities that our students do (lecture tutorials, think-pair-share questions, ranking tasks, etc.). However, one has to consider how much course time can be devoted to a course theme versus the rest of the topics that need to be covered, and this would limit the number of theme-related concepts that could be explored.

In the future, when such essays are offered, if more class time is to be spent on concepts and how to incorporate them into essays, it may be better to offer only two essays instead of three, and have all students do both. Fewer additional in-class activities will be needed and regular coverage of course topics will be minimally impacted if only two essays are offered. Another advantage is that students can be given more than two weeks to work on an essay. Pushing back the deadline will ensure sufficient time for relevant in-class activities prior to submission.

We also recognize that we unintentionally made Essay 3 more difficult than Essays 1 and 2. (Essays 1 and 2 are probably comparable in difficulty.) In the future, we will attempt to keep the difficulty level of the essays comparable by keeping around $20 - 25\%$ of the essay score or less for conceptual questions, and not ask students to address too many of these within a single essay.

We will also allocate more of the course grade to essays so that students recognize the need to devote sufficient time and effort to them in order to earn a score of C or higher. They will feel better rewarded for the time and effort they put into such projects if they form a larger part of the course grade. Since we have mentioned that the essays could be simplified, it is not clear how much more they would be weighted in the future. We would possibly double the weighting to 5%. This would cause a reduction in the weighting of our tests.

We consider the above modifications as the most important ones for future offerings of this set of essays. In the broader context, if other astronomy course themes were adopted for conceptually oriented essays, we would accordingly develop in-class activities or use existing activities for concepts similar to those in Essays 1 and 3. We would help students with writing about concepts similar to those in Essay 2. We would likely offer two essays only and ensure that the number of concepts they cover not overwhelm our students.

Other changes that would be easy to implement and possibly improve student success with essays would be:

a) to make sample essays available so students have an idea of what is expected for an A or a B grade. In the past, with group projects, we have shown students a high quality and a low quality assignment and asked them which one they would assign the higher grade to. This helps them to be aware of what to include and what to avoid in a project.

b) to remind students about the details expected and how these are connected to the grading rubrics, as suggested by one of our students.

c) to ask JJC's writing and reading center's tutors to visit our introductory astronomy classes when essays are offered to provide guidance and possibly a sample essay. In Spring 2013 and Fall 2013, we asked JJC's writing and reading center director to visit our class to provide guidance on writing essays to our Life in the Universe students. The director also had one of the center tutors create a sample that was given to all students.

d) to modify our handouts for each essay so fewer students feel confused and overwhelmed by them. We may make them more closely aligned with the essay guidelines suggested by JJC's writing and reading center, or those used in English classes.

Transits, occultations and eclipses will continue to be the news, and we would like to cover these through lecture and essays (or suitable projects) in the future as we have done here. The next two opportunities related to the theme presented here will be when the transit of Mercury occurs on 9 May 2016 (http://eclipse.gsfc.nasa.gov/transit/transit.html) and when the total solar eclipse of 21 Aug 2017 occurs (http://www.mreclipse.com/Special/SEnext.html; http://eclipse.gsfc.nasa.gov/SEplot/SEplot2001/SE2017Aug21T.GIF). The former will be entirely visible from Joliet, IL from around 6:12am till 1:42pm local time, weather permitting (http://eclipse.gsfc.nasa.gov/transit/transit.html). A telescope with a solar filter will be needed to see the event. The latter will be visible (weather permitting again) from Illinois and other parts of North America. The eclipse will have the greatest duration close to Hopkinsville, IL, with the Sun being completely blocked for 2 minutes 40.1 seconds, at this location. This will occur around 1:15pm on 21 Aug 2017 (http://eclipse.gsfc.nasa.gov/SEplot/SEplot2001/SE2017Aug21T.GIF). Solar filters will be needed to view the eclipse safely.

In the broader context, we would like to have other course themes focusing on upcoming or ongoing astronomical events or on some broad aspect of the course like stars, galaxies, etc. so we can help non-science major students to learn some of the associated concepts more deeply. We feel that devoting class time to such a theme is worthwhile, provided there is sufficient time to cover the rest of the course material with minimal impact, and if a majority of students feel that writing such essays is educationally beneficial.

## 8.    Conclusions

Through theme-related questions on one test and the final exam, we found that students averaged 63% when asked what is meant by a transit, an occultation and the "transit of Venus." Since this average is higher than the 52% we predicted in Section 2, it indicates that two of our main goals, Ia and IIa, were met through lecture.

The test and final exam questions showed that students' understanding of how an object's angular size is affected by its distance from Earth improved as the semester progressed. They also did well on questions related to how rotation period affects the duration of a transit and how much of a background object is blocked when the distance of the foreground object is changed in a transit.

We also used short theme-related essays to help students further explore the theme and related concepts. The essays were meant to be more of a teaching tool than primarily an information gathering exercise for students. The essays showed that students scored at least 70% when explaining what is a transit, an occultation and an eclipse and the need for two objects and an observer or spacecraft to be in a straight line for these events to occur. This means that our second major goal, IIa, was met through essays, in addition to lecture.

Students' understanding that the rotation period of an object does not affect a transit improved from Essay 1 to Essay 3. However, the "stand alone" transit concepts in Essay 3 were harder to grasp than the concepts in Essay 1. Students did well with the more descriptive parts of the essays.

Students felt they needed more guidance and preparation in order to write successful essays. In the future, we will provide the necessary scaffolding, such as in-class activities and online quizzes, to help students be more successful with the "stand-alone" conceptual portions of essays, as those were what students found to be the most difficult. We will also allocate around $20 - 25\%$ or less of each essay score to deep conceptual learning, and limit the number of such concepts in the future. This will ensure that the essays are not so difficult that they overwhelm students, and will increase the chance for student success.

The third major goal, to increase student interest in astronomy through short essays, did not meet with success. 21% of students felt that the essays increased their interest in astronomy. 28% of students felt that the educational value of the essays was such that they could be offered again, while 49% disagreed.

The fraction of students whose interest in astronomy increased due to these essays was essentially independent of their average essay score. This indicates that it may not be easy to increase interest in astronomy through the set of essays presented here.

40% of students who had average essay scores in the A, B or C range viewed the essays as being educationally beneficial and recommended that they be offered again. However, only 20% of those whose essays earned Ds and none of those whose essays earned Fs felt positively towards the essays. If more students earned As, Bs or Cs on their essays, then it is possible that a majority of students would view them as educationally beneficial and as worthwhile assignments for future students.

In conclusion, we feel that using short thematic essays in the introductory astronomy college course as a teaching tool is worthwhile provided a majority of students feel they are educationally beneficial. We hope that future essay offerings and associated activities will be more rewarding to students, and will motivate more students to be interested in astronomy.

## Acknowledgements

We thank our Spring 2012 face-to-face ASTR 101 students for generously providing feedback on the essay assignments, and for agreeing to let us use their essays and test answers in this project. We are very grateful to Tim Slater for reading an earlier draft and providing valuable comments that have improved the paper. We thank Vikram Dwarkadas, whose comments on a subsequent draft improved the paper too. We are also very grateful to the valuable comments and suggestions of the two anonymous referees, which have resulted in a much better version of this paper.

SLATER, T. F.; ADAMS, J. P. **Learner Centered Astronomy Teaching**: Strategies for ASTRO 101. Upper Saddle River, NJ: Pearson, 2003. 167p.

WINN, J. N. Exoplanet Transits and Occultations. In: SEAGER, S. (Ed.) **Exoplanets**. Tucson, AZ: University of Arizona Press, 2011, p. 55-77.

**APPENDIX A: Grading essays for English language skills versus science content.**

The science content expected to be in each essay has been detailed in Section 3. The English language skills that were graded for each essay were:

1) a well-written introduction
2) a well-written conclusion
3) spelling, punctuation, grammar, ease of reading (or flow), font size, text spacing, essay length, formatting (is the text broken into reasonably sized paragraphs?) were graded together.

About one-third of each essay's grade was devoted to English language skills.

The main error with writing an introduction was that the introductory paragraph would start either abruptly or with a very short "introduction" (one or two sentences) followed by the main essay content. This means the introduction is essentially missing.

The most frequent error with the conclusion was that either the concluding paragraph would be missing or it would be so short it would not sufficiently serve as a conclusion to the essay.

Both these errors were easy to identify and separate from the scientific content. If there were content errors in either introduction or conclusion, then these was graded as part of the content. The chance that a student had points deducted twice (i.e. for science content and English language) for the same error in the introduction and conclusion was very small.

Spelling, punctuation, font size, text spacing, essay length, formatting errors can be easily identified as separate errors from content, so there was no ambiguity in grading these.

The content asked for was specific, so errors in flow and grammar were not easily confused with errors in content. The flow of the writing can affect the content presentation, but choppiness in the writing is more of a distraction to the reader than a problem with identifying what message the writing attempts to convey. Similarly, it was usually clear whether or not the required content was present and satisfactorily addressed even if there were grammatical errors.

If confusing or unclear sentences were in the essay, these were usually graded as content related.

One point (out of 30 – 35 total points for each essay) was usually deducted for a combination of three errors associated with spelling, punctuation and grammar.

Font size and text spacing are related to essay length, so one point (out of 30 – 35 total points for each essay) was deducted for an essay that was around half a page too short. An essay with multiple unusually long paragraphs had 1 point deducted for this error.

Based on the above, it is unlikely points were deducted twice for the same error. Therefore it is unlikely that essay scores were lower than they should have been because both science content and English language skills were graded.

The average scores for English language skills in Essays 2 and 3 were higher, but not much higher than the average scores for these essays, indicating that the science content of these essays was somewhat more challenging for these students than their English language skills. The average essay scores for Essay 1 and its English skills were comparable.